\documentclass[final,english]{bullsrsl}[2022/06/15]

\usepackage[latin1]{inputenc}
\usepackage[T1]{fontenc}

\usepackage[numbers]{natbib} % Add "numbers" option if natbib is used
\usepackage{graphicx}
\newcommand{\kms}{\ensuremath{{\rm km}\,{\rm s}^{-1}}}

\newcommand{\msol}{M$_\odot$}

\newcommand{\teff}{T$_{\rm eff}$}
\newcommand{\logg}{$\log{g}$}
\newcommand{\porb}{$P_{\rm orb}$}

\begin{document}
\title{Detailed follow up studies of three ultracompact sdB binaries}

\author[affil={1},corresponding=true]{Eric}{Stringer}
\author[affil={1}]{Thomas}{Kupfer}
\author[affil={2}]{Matti}{Dorsch}
\affiliation[1]{Department of Physics and Astronomy, Texas Tech University, PO Box 41051, Lubbock, TX 79409, USA}
\affiliation[2]{Dr. Karl Remeis-Observatory \& ECAP, Friedrich-Alexander University Erlangen-N\"urnberg, Sternwartstr. 7, 96049 Bamberg, Germany}

\correspondance{erstring@ttu.edu}
\date{13th October 2020}
\maketitle

% \author[affil1]{FirstName (+ MiddleInitials if necessary)}{FamilyName}
% \author[affil2]{...}{}
% \equalcontribauthor[]{}{} % Maximum two --> counter
% \consortium[affil]{Consortium Name}
% With consortium: affiliation will be set to "See Appendix 1 for a full
% list of consortium members and their respective affiliations
% \affiliation[affil1]{...}
% \affiliationq[affil2]{...}

% \correspondence[]{}
% No explicit corresponding author: use first author
% 

% Abstract of the paper in the same language as the paper
\begin{abstract}
We present follow-up studies of three ultracompact hot subdwarf binaries. Using data from the Zwicky Transient Facility, we find orbital periods of 33.6, 37.3, and 36.9 minutes for ZTF\,1946+3203, ZTF\,0640+1738, and ZTF\,0643+0318 respectively. The light curves show ellipsoidal variability of the hot subdwarf star with potential eclipses of an accretion disc. Phase-resolved spectroscopic observations with Keck were used to measure a radial velocity curve and atmospheric parameters of the hot subdwarf stars. ZTF\,J0643 shows evidence of accretion disc emission lines in the average spectrum. Combining light curve and spectroscopic fits will allow us to measure precise system properties such as masses, to determine the evolutionary history and future evolution of the system.
%Then the spectra from Keck were used to find the radial velocities and atmospheric properties of each binary. Our goal is to measure the system properties such as masses, to determine the evolutionary history and future of the system and confirm them as type Ia supernova progenitors.
\end{abstract}

\section{Introduction}
Most hot subdwarf B stars (sdBs) are core helium burning stars with masses around 0.5 \msol\, and thin hydrogen envelopes \citep{heb86,heb09,heb16}. A large number of sdB stars are in close orbits with orbital periods of \porb$<10$\,days \citep{nap04a,max01}, with the most compact systems reaching orbital periods of $<1$\,hour \citep{ven12,gei13,kup17,kup17a, kup20, kup20a}.

SdB binaries with white dwarf (WD) companions which exit the common envelope phase at \porb$<2$\,hours will reach contact while the sdB is still burning helium \citep{bau21}. Once the sdB fills its Roche Lobe, helium-rich material will be transferred to the WD companion. After the WD accretes $\approx$0.1\,\msol, helium burning is predicted to be ignited unstably in the accreted helium layer on the WD surface \citep{bro15,bau17}. This could either disrupt the WD even when the mass is significantly below the Chandrasekhar mass, a so-called double detonation supernova \citep{liv90,liv95,fin10,woo11,wan12,she14,wan18} or just detonate the He-shell without disrupting the WD which results in a faint and fast .Ia supernova with subsequent weaker He-flashes \citep{bil07,bro15}.

In this work we present follow-up studies of three ultracompact hot subdwarf binaries which were first discovered by Burdge et al. 2020 \citep{bur20}. The objects are ZTF\,1946+3203 (ZTF\,J1946), ZTF\,0640+1738 (ZTF\,J0640), and ZTF\,0643+0318 (ZTF\,J0643). They show periods of 34, 37, and 37 minutes based on their photometric variability, making them the most compact hot subdwarf binaries known today. As such they are ideal candidates to be type Ia supernova progenitors. One of the systems has signs of an accretion disc, making it the first candidate with spectral signatures for an accretion disk.

\begin{figure}
\centering
\label{fig:sum}
\includegraphics[scale=0.8]{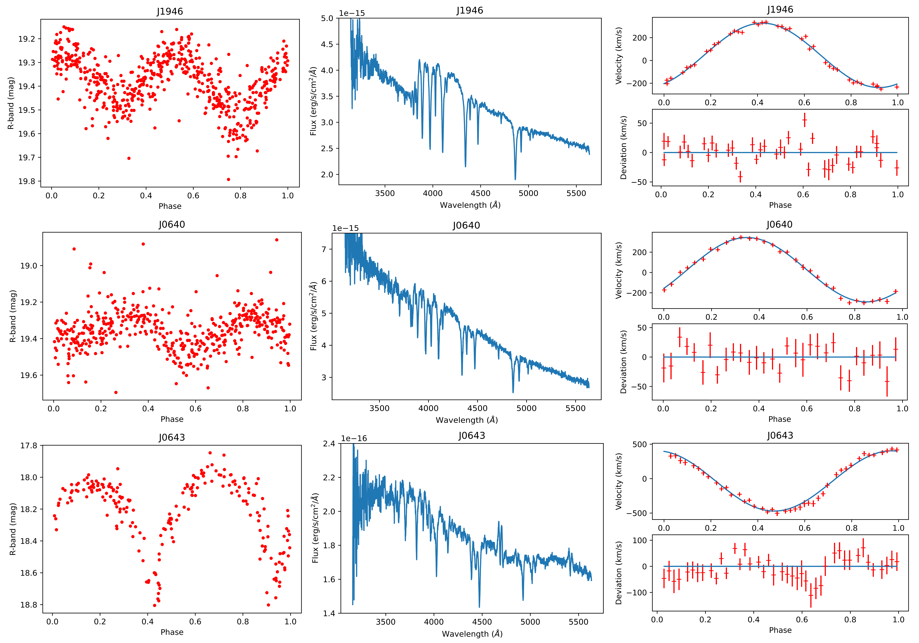}
\bigskip

\begin{minipage}{12cm}
\caption{Left Column: Phase folded ZTF light curves. Middle Column: Spectra from Keck LRIS with 600/4000 grating. Right Column:(Top) Velocity curves with accompanying sine-fit. (Bottom) Residuals of the fit}
\end{minipage}
\end{figure}

%Type Ia supernovae have traditionally been important in determining the distances of galaxies. Compact hot subdwarf binaries are potentially the progenitors of a sub-type of type Ia supernovae, where the hot subdwarfs transfers helium-rich material to the white dwarf companion leading to a double detonation supernova.

%Hot subdwarf binaries with white dwarf companions that formed at orbital periods below approximately two hours can reach contact before the hot subdwarf turns into a white dwarf. Only recently the first members of this class of very compact sdB systems has been found, and less than a handful are known. The most compact system known so far is ZTFJ2130 with an orbital period of 39 minutes.

\section{Photometric analysis}
All three objects were observed as part of the Zwicky Transient Facility (ZTF) public survey \citep{gra19,bel19}. Image processing of ZTF data is described in full detail in Masci et al. 2019 \citep{mas19}. The number of ZTF epochs range from 413 to 1004 points over three years for each object. To measure the orbital period we performed a Lomb-Scargle analysis on each data set using the Astropy Lomb-Scargle periodogram module\footnote{https://docs.astropy.org/en/stable/timeseries/lombscargle.html}\citep{astpy13, astpy18}. 

We find orbital periods of 33.6\,min, 37.3\,min, and 36.9\,min for ZTF\,J1946, ZTF\,0640, and ZTF\,0643 respectively. This is in agreement with previous the studies \citep{bur20}. The phase folded light curves show strong periodic ellipsoidal variability. ZTF\,J1946 shows a deep eclipse which is also seen in the known sdOBs with a confirmed accretion disc \citep{kup20, kup20a}. ZTF\,J0643 shows equally deep eclipses leading to the conclusion that the second component, most likely the accretion disc, has a similar brightness than the hot subdwarf primary (see left panels in Fig.\,\ref{fig:sum}).

\section{Spectroscopic analysis}
Phase resolved spectroscopy was taken with Keck and the Low Resolution Imaging Spectrometer (LRIS; \citep{oke95}). ZTF\,J1946, ZTF\,J0640, and ZTF\,J0643 has 31, 40, and 42 phase resolved spectra respectively covering a full orbit for each objects. To measure the radial velocities, we used the cross correlation \texttt{rvsao} package implemented in \texttt{IRAF}. We then performed a sine-fit to the individual radial velocities to measure the radial velocity semi-amplitude for each object. We find a radial velocity semi-amplitudes of $282.6\pm18.7$\,\kms, $316.7\pm26.8$\,\kms\, and $437.4\pm84.7$\,\kms\, for ZTF\,J1946, ZTF\,J0640, and ZTF\,J0643 respectively (see right panels in Fig.\,\ref{fig:sum}). The large error for the radial velocity semi-amplitude in ZTF\,J0643 is likely originating from the contamination from the accretion disc. 

%the spectra for each object were summed and cross-correlated with IRAF. 

The average spectra were constructed from the sum of the individual rest-wavelength corrected spectra. ZTF\,J1946 and ZTF\,J0640 show typical hot subdwarf spectra with strong hydrogen and helium absorption features. ZTF\,J0643 shows a spectrum of a He-sdO with strong helium and weak hydrogen absorption lines as well as double peaked emission lines in He\,{\sc ii} 4686\,\AA\, and He\,{\sc ii} 5411\,\AA\, indicating emission from an accretion disc (see middle panels in Fig.\,\ref{fig:sum}). This makes ZTF\,J0643 the first hot subdwarf binary with spectral signatures of an accretion disc. 

We use spectral models to fit the co-added spectra of ZTF\,J1946 and ZTF\,J0640 which were constructed using a hybrid LTE/NLTE approach described in detail in Przybilla et al. 2011 \citep{prz11} and Irrgang et al. 2021 \citep{irr21}. The grid of spectral models covers a typical range of hot subdwarf \teff\, and \logg\, up to modest helium abundances (Heber et al. submitted). ZTF\,J0643 shows strong helium features and requires larger helium abundances. Therefore, for ZTF\,J0643 we use spectral models computed with TLUSTY/SYNSPEC \citep{hub17} covering helium dominated atmospheres. The approach and the models are described in detail in Dorsch et al. 2022 \citep{dor22}. Using the spectral modelling tool SPAS \citep{hir09}, the co-added spectra were fitted for effective temperature (\teff), surface gravity (\logg), and helium abundance ($\log(y)=\log\frac{n(He)}{n(H)}$) using atmospheric models. For ZTF\,J1946, we find \teff$=27,500\pm1000$\,K, \logg$=5.90\pm0.10$, and $\log(y)=-1.11\pm0.10$. For ZTF\,J0640 we find \teff$=30,500\pm1000$\,K, \logg$=5.70\,\pm0.15$, and $\log(y)=-0.38\pm0.20$. Finally for ZTF\,J0643 we find \teff$=42,500\pm2000$\,K, \logg$=6.15\,\pm0.30$, and $\log(y)=+1.8_{-0.3}^{+\infty}$.

\section{Conclusion and summary}
We present detailed follow-up observations of three new ultracompact hot subdwarf binaries with orbital periods below 40min, making them the most compact hot subdwarfs known today. ZTF\,J0640 shows ellipsoidal deformation whereas ZTF\,J1946 and ZTF\,J0643 show light curve shapes typical for mass transferring systems with a deep eclipse on top of ellipsoidal deformation indicating the presence of an accretion disc. ZTF\,J0643 presents double peaked helium emission features from an accretion disc providing the first direct evidence for an accretion disc in a hot subdwarf binary as well as a low-hydrogen content. The low hydrogen content suggests that the system has lost a large fraction of its hydrogen envelope and potentially shows an advanced stage of mass-transferring hot subdwarfs \citep{bau21}. 

To continue this study, we will create light curve models using the light curve modelling code LCURVE \citep{cop10} with prior information from spectroscopy, and calculate system properties such as masses, radii and inclination angle to be able to fully constrain the evolutionary history of each system.

% After your conclusions, you may at your convenience include an
% "Acknowledgments" section with
\begin{acknowledgments}
TK acknowledges support from the National Science Foundation through grant AST \#2107982, from NASA through grant 80NSSC22K0338 and from STScI through grant HST-GO-16659.002-A. We thank Alekzander Koaskowski for feedback on the manuscript.

Based on observations obtained with the Samuel Oschin 48-inch Telescope at the Palomar Observatory as part of the Zwicky Transient Facility project. ZTF is supported by the National Science Foundation under Grant No. AST-1440341 and a collaboration including Caltech, IPAC, the Weizmann Institute for Science, the Oskar Klein Center at Stockholm University, the University of Maryland, the University of Washington, Deutsches Elektronen-Synchrotron and Humboldt University, Los Alamos National Laboratories, the TANGO Consortium of Taiwan, the University of Wisconsin at Milwaukee, and Lawrence Berkeley National Laboratories. Operations are conducted by COO, IPAC, and UW.

Some of the data presented herein were obtained at the W.M. Keck Observatory, which is operated as a scientific partnership among the California Institute of Technology, the University of California and the National Aeronautics and Space Administration. The Observatory was made possible by the generous financial support of the W.M. Keck Foundation. The authors wish to recognize and acknowledge the very significant cultural role and reverence that the summit of Mauna Kea has always had within the indigenous Hawaiian community. We are most fortunate to have the opportunity to conduct observations from this mountain. 

\end{acknowledgments}
% If you prefer not to include any acknowledgments, please simply omit the
% complete environment.

% Finally, there are several bits and pieces of information that may,
% or that have to be included. These go into the "Further information"
% section (which is mandatory).

\begin{furtherinformation}

% First is the optional "orcids" section. It can be used to list the
% ORCIDs of those authors that would like to share them.
\begin{orcids}
% Please list them, one per line, with the command
% \orcid{ORCID}{First name}{Last name}.
%
% Example:
% \orcid{0000-1111-2222-3333}{Hàrry}{Harrisòn}
% \orcid{1111-2222-3333-4444}{Leonie}{van Leon}
% \orcid{2222-3333-4444-5555}{Lotta}{Lothardis}
\orcid{0000-0002-6540-1484}{Thomas}{Kupfer}
%\orcid{...}{First name + Middle Initials if any}{Last Name}
%\orcid{...}{First name + Middle Initials if any}{Last Name}
%....
\end{orcids}
% If none of the author wants to include their ORCID, please leave out
% (comment out or delete) the "orcids" environment

% Next comes the "Author contributions" section, which is mandatory for
% mansucripts signed by more than one author. The contributions of each
% author (identified by their initials) must be declared.
% We recommend the CRediT taxonomy \href{http://credit.niso.org}{CRediT}
% (Contributor Roles Taxonomy).
% If there is only one author, this section should be omitted
% (commented out or deleted, not left empty).
\begin{authorcontributions}
ES extracted the ZTF light curves, performed the period analysis and radial velocity measurements, supported the interpretation of the spectra as well as led the writing of the manuscript. TK performed the spectral modelling with SPAS and supported the writing of this manuscript. MD supported the spectral fitting and some of the spectral models used in this work.
\end{authorcontributions}

% Finally, the "Conflicts of interest" declaraition is andatory for every
% paper (with single or multiple authors, ...) , even if there is no conflict
% of interest. Authors must declare any personal or professional circumstances
% that may be perceived as influencing the results or the conlusions reported
% in the manuscript.
\begin{conflictsofinterest}
% If there is no conflict of interest, please state that
% ``The authors declare no conflict of interest.''
The authors declare no conflict of interest.
\end{conflictsofinterest}

\end{furtherinformation}

% References
% ==========
% References can be listed alphabetically by the first author's last name, with
% in-text citation in (Author, Year) style. Alternatively, they can also be
% numbered in order of appearance in the text [$n$].
% All common formats of references are accepted, but consistency throughout the
% reference list is required.

% We recommend to use the "natbib" package, which allows to easily switch
% between the numbered and the author-year based styles, simply by requesting
% the adequate bibliography style with the \bibliographystyle{...} command,
% and adding (resp.\ removing) the "numbers" option with the
% \usepackage{natbib} declaration. With numbered references, it is also possible
% to use the standard LaTeX \cite commands.

% bullsrsl.cls is completed by four BibTeX style files. Two of them are for
% manuscripts written in English:
%  - bullsrsl-en.bst: for author-year citations in English manuscripts
%    (requires the natbib package);
%  - bullsrsl-numen.bst: for numbered citations in English manuscripts
%    (may be used with or without natbib);
%
% Make sure to include DOIs whenever they are available for all the cited
% references.

\bibliographystyle{bullsrsl-numen}
%\bibliographystyle{bullsrsl-en}

% If you use BibTeX, please list the bibliography databases in the
% \bibliography{...} command. Please notice that you will have to provide us
% with a (sub-setted) single file database once your manuscript has been
% accepted for publication and goes into production. Sub-setting can be easily
% done with bibexport.sh or bibtool, both available from CTAN.
\bibliography{refs, refs_1508}

\begin{thebibliography}{10}
\providecommand{\url}[1]{#1}
\providecommand{\urlprefix}{URL }

\bibitem{heb86}
{Heber}, U. (1986) {The atmosphere of subluminous B stars. II - Analysis of 10
  helium poor subdwarfs and the birthrate of sdB stars}.
\newblock A\&A, 155, 33--45.

\bibitem{heb09}
{Heber}, U. (2009) {Hot Subdwarf Stars}.
\newblock ARA\&A, 47, 211--251.
\newblock \url{https://doi.org/10.1146/annurev-astro-082708-101836}.

\bibitem{heb16}
{Heber}, U. (2016) {Hot Subluminous Stars}.
\newblock PASP, 128(8), 082001.
\newblock \url{https://doi.org/10.1088/1538-3873/128/966/082001}.

\bibitem{nap04a}
{Napiwotzki}, R., {Karl}, C.~A., {Lisker}, T., {Heber}, U., {Christlieb}, N.,
  {Reimers}, D., {Nelemans}, G. and {Homeier}, D. (2004) {Close binary EHB
  stars from SPY}.
\newblock Astrophysics and Space Science, 291, 321--328.
\newblock \url{https://doi.org/10.1023/B:ASTR.0000044362.07416.6c}.

\bibitem{max01}
{Maxted}, P.~f.~L., {Heber}, U., {Marsh}, T.~R. and {North}, R.~C. (2001) {The
  binary fraction of extreme horizontal branch stars}.
\newblock MNRAS, 326, 1391--1402.
\newblock \url{https://doi.org/10.1111/j.1365-8711.2001.04714.x}.

\bibitem{ven12}
{Vennes}, S., {Kawka}, A., {O'Toole}, S.~J., {N{\'e}meth}, P. and {Burton}, D.
  (2012) {The Shortest Period sdB Plus White Dwarf Binary CD-30 11223 (GALEX
  J1411-3053)}.
\newblock ApJL, 759, L25.
\newblock \url{https://doi.org/10.1088/2041-8205/759/1/L25}.

\bibitem{gei13}
{Geier}, S., {Marsh}, T.~R., {Wang}, B., {Dunlap}, B., {Barlow}, B.~N.,
  {Schaffenroth}, V., {Chen}, X., {Irrgang}, A., {Maxted}, P.~F.~L.,
  {Ziegerer}, E., {Kupfer}, T., {Miszalski}, B., {Heber}, U., {Han}, Z.,
  {Shporer}, A., {Telting}, J.~H., {G{\"a}nsicke}, B.~T., {{\O}stensen}, R.~H.,
  {O'Toole}, S.~J. and {Napiwotzki}, R. (2013) {A progenitor binary and an
  ejected mass donor remnant of faint type Ia supernovae}.
\newblock A\&A, 554, A54.
\newblock \url{https://doi.org/10.1051/0004-6361/201321395}.

\bibitem{kup17}
{Kupfer}, T., {van Roestel}, J., {Brooks}, J., {Geier}, S., {Marsh}, T.~R.,
  {Groot}, P.~J., {Bloemen}, S., {Prince}, T.~A., {Bellm}, E., {Heber}, U.,
  {Bildsten}, L., {Miller}, A.~A., {Dyer}, M.~J., {Dhillon}, V.~S., {Green},
  M., {Irawati}, P., {Laher}, R., {Littlefair}, S.~P., {Shupe}, D.~L.,
  {Steidel}, C.~C., {Rattansoon}, S. and {Pettini}, M. (2017) {PTF1
  J082340.04+081936.5: A Hot Subdwarf B Star with a Low-mass White Dwarf
  Companion in an 87-minute Orbit}.
\newblock ApJ, 835, 131.
\newblock \url{https://doi.org/10.3847/1538-4357/835/2/131}.

\bibitem{kup17a}
{Kupfer}, T., {Ramsay}, G., {van Roestel}, J., {Brooks}, J., {MacFarlane},
  S.~A., {Toma}, R., {Groot}, P.~J., {Woudt}, P.~A., {Bildsten}, L., {Marsh},
  T.~R., {Green}, M.~J., {Breedt}, E., {Kilkenny}, D., {Freudenthal}, J.,
  {Geier}, S., {Heber}, U., {Bagnulo}, S., {Blagorodnova}, N., {Buckley},
  D.~A.~H., {Dhillon}, V.~S., {Kulkarni}, S.~R., {Lunnan}, R. and {Prince},
  T.~A. (2017) {The OmegaWhite Survey for Short-period Variable Stars. V.
  Discovery of an Ultracompact Hot Subdwarf Binary with a Compact Companion in
  a 44-minute Orbit}.
\newblock ApJ, 851, 28.
\newblock \url{https://doi.org/10.3847/1538-4357/aa9522}.

\bibitem{kup20}
{Kupfer}, T., {Bauer}, E.~B., {Marsh}, T.~R., {van Roestel}, J., {Bellm},
  E.~C., {Burdge}, K.~B., {Coughlin}, M.~W., {Fuller}, J., {Hermes}, J.,
  {Bildsten}, L., {Kulkarni}, S.~R., {Prince}, T.~A., {Szkody}, P., {Dhillon},
  V.~S., {Murawski}, G., {Burruss}, R., {Dekany}, R., {Delacroix}, A., {Drake},
  A.~J., {Duev}, D.~A., {Feeney}, M., {Graham}, M.~J., {Kaplan}, D.~L.,
  {Laher}, R.~R., {Littlefair}, S.~P., {Masci}, F.~J., {Riddle}, R.,
  {Rusholme}, B., {Serabyn}, E., {Smith}, R.~M., {Shupe}, D.~L. and
  {Soumagnac}, M.~T. (2020) {The First Ultracompact Roche Lobe-Filling Hot
  Subdwarf Binary}.
\newblock ApJ, 891(1), 45.
\newblock \url{https://doi.org/10.3847/1538-4357/ab72ff}.

\bibitem{kup20a}
{Kupfer}, T., {Bauer}, E.~B., {Burdge}, K.~B., {Roestel}, J.~v., {Bellm},
  E.~C., {Fuller}, J., {Hermes}, J., {Marsh}, T.~R., {Bildsten}, L.,
  {Kulkarni}, S.~R., {Phinney}, E.~S., {Prince}, T.~A., {Szkody}, P., {Yao},
  Y., {Irrgang}, A., {Heber}, U., {Schneider}, D., {Dhillon}, V.~S.,
  {Murawski}, G., {Drake}, A.~J., {Duev}, D.~A., {Feeney}, M., {Graham}, M.~J.,
  {Laher}, R.~R., {Littlefair}, S.~P., {Mahabal}, A.~A., {Masci}, F.~J.,
  {Porter}, M., {Reiley}, D., {Rodriguez}, H., {Rusholme}, B., {Shupe}, D.~L.
  and {Soumagnac}, M.~T. (2020) {A New Class of Roche Lobe-filling Hot Subdwarf
  Binaries}.
\newblock ApJL, 898(1), L25.
\newblock \url{https://doi.org/10.3847/2041-8213/aba3c2}.

\bibitem{bau21}
{Bauer}, E.~B. and {Kupfer}, T. (2021) {Phases of Mass Transfer from Hot
  Subdwarfs to White Dwarf Companions and Their Photometric Properties}.
\newblock ApJ, 922(2), 245.
\newblock \url{https://doi.org/10.3847/1538-4357/ac25f0}.

\bibitem{bro15}
{Brooks}, J., {Bildsten}, L., {Marchant}, P. and {Paxton}, B. (2015) {AM Canum
  Venaticorum Progenitors with Helium Star Donors and the Resultant
  Explosions}.
\newblock ApJ, 807, 74.
\newblock \url{https://doi.org/10.1088/0004-637X/807/1/74}.

\bibitem{bau17}
{Bauer}, E.~B., {Schwab}, J. and {Bildsten}, L. (2017) {Electron Captures on
  $^{14}$N as a Trigger for Helium Shell Detonations}.
\newblock ApJ, 845, 97.
\newblock \url{https://doi.org/10.3847/1538-4357/aa7ffa}.

\bibitem{liv90}
{Livne}, E. (1990) {Successive detonations in accreting white dwarfs as an
  alternative mechanism for type I supernovae}.
\newblock ApJl, 354, L53--L55.
\newblock \url{https://doi.org/10.1086/185721}.

\bibitem{liv95}
{Livne}, E. and {Arnett}, D. (1995) {Explosions of Sub--Chandrasekhar Mass
  White Dwarfs in Two Dimensions}.
\newblock ApJ, 452, 62.
\newblock \url{https://doi.org/10.1086/176279}.

\bibitem{fin10}
{Fink}, M., {R{\"o}pke}, F.~K., {Hillebrandt}, W., {Seitenzahl}, I.~R., {Sim},
  S.~A. and {Kromer}, M. (2010) {Double-detonation sub-Chandrasekhar
  supernovae: can minimum helium shell masses detonate the core?}
\newblock A\&A, 514, A53.
\newblock \url{https://doi.org/10.1051/0004-6361/200913892}.

\bibitem{woo11}
{Woosley}, S.~E. and {Kasen}, D. (2011) {Sub-Chandrasekhar Mass Models for
  Supernovae}.
\newblock ApJ, 734, 38.
\newblock \url{https://doi.org/10.1088/0004-637X/734/1/38}.

\bibitem{wan12}
{Wang}, B. and {Han}, Z. (2012) {Progenitors of type Ia supernovae}.
\newblock New Astronomy Reviews, 56, 122--141.
\newblock \url{https://doi.org/10.1016/j.newar.2012.04.001}.

\bibitem{she14}
{Shen}, K.~J. and {Bildsten}, L. (2014) {The Ignition of Carbon Detonations via
  Converging Shock Waves in White Dwarfs}.
\newblock ApJ, 785, 61.
\newblock \url{https://doi.org/10.1088/0004-637X/785/1/61}.

\bibitem{wan18}
{Wang}, B. (2018) {Mass-accreting white dwarfs and type Ia supernovae}.
\newblock Research in Astronomy and Astrophysics, 18(5), 049.
\newblock \url{https://doi.org/10.1088/1674-4527/18/5/49}.

\bibitem{bil07}
{Bildsten}, L., {Shen}, K.~J., {Weinberg}, N.~N. and {Nelemans}, G. (2007)
  {Faint Thermonuclear Supernovae from AM Canum Venaticorum Binaries}.
\newblock ApJL, 662, L95--L98.
\newblock \url{https://doi.org/10.1086/519489}.

\bibitem{bur20}
{Burdge}, K.~B., {Prince}, T.~A., {Fuller}, J., {Kaplan}, D.~L., {Marsh},
  T.~R., {Tremblay}, P.-E., {Zhuang}, Z., {Bellm}, E.~C., {Caiazzo}, I.,
  {Coughlin}, M.~W., {Dhillon}, V.~S., {Gaensicke}, B., {Rodr{\'\i}guez-Gil},
  P., {Graham}, M.~J., {Hermes}, J., {Kupfer}, T., {Littlefair}, S.~P.,
  {Mr{\'o}z}, P., {Phinney}, E.~S., {van Roestel}, J., {Yao}, Y., {Dekany},
  R.~G., {Drake}, A.~J., {Duev}, D.~A., {Hale}, D., {Feeney}, M., {Helou}, G.,
  {Kaye}, S., {Mahabal}, A.~A., {Masci}, F.~J., {Riddle}, R., {Smith}, R.,
  {Soumagnac}, M.~T. and {Kulkarni}, S.~R. (2020) {A Systematic Search of
  Zwicky Transient Facility Data for Ultracompact Binary LISA-detectable
  Gravitational-wave Sources}.
\newblock ApJ, 905(1), 32.
\newblock \url{https://doi.org/10.3847/1538-4357/abc261}.

\bibitem{gra19}
{Graham}, M.~J., {Kulkarni}, S.~R., {Bellm}, E.~C., {Adams}, S.~M.,
  {Barbarino}, C., {Blagorodnova}, N., {Bodewits}, D., {Bolin}, B., {Brady},
  P.~R., {Cenko}, S.~B., {Chang}, C.-K., {Coughlin}, M.~W., {De}, K., {Eadie},
  G., {Farnham}, T.~L., {Feindt}, U., {Franckowiak}, A., {Fremling}, C.,
  {Gezari}, S., {Ghosh}, S., {Goldstein}, D.~A., {Golkhou}, V.~Z., {Goobar},
  A., {Ho}, A. Y.~Q., {Huppenkothen}, D., {Ivezi{\'c}}, {\v{Z}}., {Jones},
  R.~L., {Juric}, M., {Kaplan}, D.~L., {Kasliwal}, M.~M., {Kelley}, M. S.~P.,
  {Kupfer}, T., {Lee}, C.-D., {Lin}, H.~W., {Lunnan}, R., {Mahabal}, A.~A.,
  {Miller}, A.~A., {Ngeow}, C.-C., {Nugent}, P., {Ofek}, E.~O., {Prince},
  T.~A., {Rauch}, L., {van Roestel}, J., {Schulze}, S., {Singer}, L.~P.,
  {Sollerman}, J., {Taddia}, F., {Yan}, L., {Ye}, Q.-Z., {Yu}, P.-C., {Barlow},
  T., {Bauer}, J., {Beck}, R., {Belicki}, J., {Biswas}, R., {Brinnel}, V.,
  {Brooke}, T., {Bue}, B., {Bulla}, M., {Burruss}, R., {Connolly}, A.,
  {Cromer}, J., {Cunningham}, V., {Dekany}, R., {Delacroix}, A., {Desai}, V.,
  {Duev}, D.~A., {Feeney}, M., {Flynn}, D., {Frederick}, S., {Gal-Yam}, A.,
  {Giomi}, M., {Groom}, S., {Hacopians}, E., {Hale}, D., {Helou}, G.,
  {Henning}, J., {Hover}, D., {Hillenbrand}, L.~A., {Howell}, J., {Hung}, T.,
  {Imel}, D., {Ip}, W.-H., {Jackson}, E., {Kaspi}, S., {Kaye}, S., {Kowalski},
  M., {Kramer}, E., {Kuhn}, M., {Landry}, W., {Laher}, R.~R., {Mao}, P.,
  {Masci}, F.~J., {Monkewitz}, S., {Murphy}, P., {Nordin}, J., {Patterson},
  M.~T., {Penprase}, B., {Porter}, M., {Rebbapragada}, U., {Reiley}, D.,
  {Riddle}, R., {Rigault}, M., {Rodriguez}, H., {Rusholme}, B., {van Santen},
  J., {Shupe}, D.~L., {Smith}, R.~M., {Soumagnac}, M.~T., {Stein}, R.,
  {Surace}, J., {Szkody}, P., {Terek}, S., {Van Sistine}, A., {van Velzen}, S.,
  {Vestrand}, W.~T., {Walters}, R., {Ward}, C., {Zhang}, C. and {Zolkower}, J.
  (2019) {The Zwicky Transient Facility: Science Objectives}.
\newblock PASP, 131(1001), 078001.
\newblock \url{https://doi.org/10.1088/1538-3873/ab006c}.

\bibitem{bel19}
{Bellm}, E.~C., {Kulkarni}, S.~R., {Graham}, M.~J., {Dekany}, R., {Smith},
  R.~M., {Riddle}, R., {Masci}, F.~J., {Helou}, G., {Prince}, T.~A., {Adams},
  S.~M., {Barbarino}, C., {Barlow}, T., {Bauer}, J., {Beck}, R., {Belicki}, J.,
  {Biswas}, R., {Blagorodnova}, N., {Bodewits}, D., {Bolin}, B., {Brinnel}, V.,
  {Brooke}, T., {Bue}, B., {Bulla}, M., {Burruss}, R., {Cenko}, S.~B., {Chang},
  C.-K., {Connolly}, A., {Coughlin}, M., {Cromer}, J., {Cunningham}, V., {De},
  K., {Delacroix}, A., {Desai}, V., {Duev}, D.~A., {Eadie}, G., {Farnham},
  T.~L., {Feeney}, M., {Feindt}, U., {Flynn}, D., {Franckowiak}, A.,
  {Frederick}, S., {Fremling}, C., {Gal-Yam}, A., {Gezari}, S., {Giomi}, M.,
  {Goldstein}, D.~A., {Golkhou}, V.~Z., {Goobar}, A., {Groom}, S., {Hacopians},
  E., {Hale}, D., {Henning}, J., {Ho}, A.~Y.~Q., {Hover}, D., {Howell}, J.,
  {Hung}, T., {Huppenkothen}, D., {Imel}, D., {Ip}, W.-H., {Ivezi{\'c}}, {\v
  Z}., {Jackson}, E., {Jones}, L., {Juric}, M., {Kasliwal}, M.~M., {Kaspi}, S.,
  {Kaye}, S., {Kelley}, M.~S.~P., {Kowalski}, M., {Kramer}, E., {Kupfer}, T.,
  {Landry}, W., {Laher}, R.~R., {Lee}, C.-D., {Lin}, H.~W., {Lin}, Z.-Y.,
  {Lunnan}, R., {Giomi}, M., {Mahabal}, A., {Mao}, P., {Miller}, A.~A.,
  {Monkewitz}, S., {Murphy}, P., {Ngeow}, C.-C., {Nordin}, J., {Nugent}, P.,
  {Ofek}, E., {Patterson}, M.~T., {Penprase}, B., {Porter}, M., {Rauch}, L.,
  {Rebbapragada}, U., {Reiley}, D., {Rigault}, M., {Rodriguez}, H., {van
  Roestel}, J., {Rusholme}, B., {van Santen}, J., {Schulze}, S., {Shupe},
  D.~L., {Singer}, L.~P., {Soumagnac}, M.~T., {Stein}, R., {Surace}, J.,
  {Sollerman}, J., {Szkody}, P., {Taddia}, F., {Terek}, S., {Van Sistine}, A.,
  {van Velzen}, S., {Vestrand}, W.~T., {Walters}, R., {Ward}, C., {Ye}, Q.-Z.,
  {Yu}, P.-C., {Yan}, L. and {Zolkower}, J. (2019) {The Zwicky Transient
  Facility: System Overview, Performance, and First Results}.
\newblock PASP, 131(1), 018002.
\newblock \url{https://doi.org/10.1088/1538-3873/aaecbe}.

\bibitem{mas19}
{Masci}, F.~J., {Laher}, R.~R., {Rusholme}, B., {Shupe}, D.~L., {Groom}, S.,
  {Surace}, J., {Jackson}, E., {Monkewitz}, S., {Beck}, R., {Flynn}, D.,
  {Terek}, S., {Landry}, W., {Hacopians}, E., {Desai}, V., {Howell}, J.,
  {Brooke}, T., {Imel}, D., {Wachter}, S., {Ye}, Q.-Z., {Lin}, H.-W., {Cenko},
  S.~B., {Cunningham}, V., {Rebbapragada}, U., {Bue}, B., {Miller}, A.~A.,
  {Mahabal}, A., {Bellm}, E.~C., {Patterson}, M.~T., {Juri{\'c}}, M.,
  {Golkhou}, V.~Z., {Ofek}, E.~O., {Walters}, R., {Graham}, M., {Kasliwal},
  M.~M., {Dekany}, R.~G., {Kupfer}, T., {Burdge}, K., {Cannella}, C.~B.,
  {Barlow}, T., {Van Sistine}, A., {Giomi}, M., {Fremling}, C., {Blagorodnova},
  N., {Levitan}, D., {Riddle}, R., {Smith}, R.~M., {Helou}, G., {Prince}, T.~A.
  and {Kulkarni}, S.~R. (2019) {The Zwicky Transient Facility: Data Processing,
  Products, and Archive}.
\newblock PASP, 131(1), 018003.
\newblock \url{https://doi.org/10.1088/1538-3873/aae8ac}.

\bibitem{astpy13}
{Astropy Collaboration}, {Robitaille}, T.~P., {Tollerud}, E.~J., {Greenfield},
  P., {Droettboom}, M., {Bray}, E., {Aldcroft}, T., {Davis}, M., {Ginsburg},
  A., {Price-Whelan}, A.~M., {Kerzendorf}, W.~E., {Conley}, A., {Crighton}, N.,
  {Barbary}, K., {Muna}, D., {Ferguson}, H., {Grollier}, F., {Parikh}, M.~M.,
  {Nair}, P.~H., {Unther}, H.~M., {Deil}, C., {Woillez}, J., {Conseil}, S.,
  {Kramer}, R., {Turner}, J.~E.~H., {Singer}, L., {Fox}, R., {Weaver}, B.~A.,
  {Zabalza}, V., {Edwards}, Z.~I., {Azalee Bostroem}, K., {Burke}, D.~J.,
  {Casey}, A.~R., {Crawford}, S.~M., {Dencheva}, N., {Ely}, J., {Jenness}, T.,
  {Labrie}, K., {Lim}, P.~L., {Pierfederici}, F., {Pontzen}, A., {Ptak}, A.,
  {Refsdal}, B., {Servillat}, M. and {Streicher}, O. (2013) {Astropy: A
  community Python package for astronomy}.
\newblock A\&A, 558, A33.
\newblock \url{https://doi.org/10.1051/0004-6361/201322068}.

\bibitem{astpy18}
{Astropy Collaboration}, {Price-Whelan}, A.~M., {Sip{\H o}cz}, B.~M.,
  {G{\"u}nther}, H.~M., {Lim}, P.~L., {Crawford}, S.~M., {Conseil}, S.,
  {Shupe}, D.~L., {Craig}, M.~W., {Dencheva}, N., {Ginsburg}, A., {VanderPlas},
  J.~T., {Bradley}, L.~D., {P{\'e}rez-Su{\'a}rez}, D., {de Val-Borro}, M.,
  {Aldcroft}, T.~L., {Cruz}, K.~L., {Robitaille}, T.~P., {Tollerud}, E.~J.,
  {Ardelean}, C., {Babej}, T., {Bach}, Y.~P., {Bachetti}, M., {Bakanov}, A.~V.,
  {Bamford}, S.~P., {Barentsen}, G., {Barmby}, P., {Baumbach}, A., {Berry},
  K.~L., {Biscani}, F., {Boquien}, M., {Bostroem}, K.~A., {Bouma}, L.~G.,
  {Brammer}, G.~B., {Bray}, E.~M., {Breytenbach}, H., {Buddelmeijer}, H.,
  {Burke}, D.~J., {Calderone}, G., {Cano Rodr{\'{\i}}guez}, J.~L., {Cara}, M.,
  {Cardoso}, J.~V.~M., {Cheedella}, S., {Copin}, Y., {Corrales}, L.,
  {Crichton}, D., {D'Avella}, D., {Deil}, C., {Depagne}, {\'E}., {Dietrich},
  J.~P., {Donath}, A., {Droettboom}, M., {Earl}, N., {Erben}, T., {Fabbro}, S.,
  {Ferreira}, L.~A., {Finethy}, T., {Fox}, R.~T., {Garrison}, L.~H., {Gibbons},
  S.~L.~J., {Goldstein}, D.~A., {Gommers}, R., {Greco}, J.~P., {Greenfield},
  P., {Groener}, A.~M., {Grollier}, F., {Hagen}, A., {Hirst}, P., {Homeier},
  D., {Horton}, A.~J., {Hosseinzadeh}, G., {Hu}, L., {Hunkeler}, J.~S.,
  {Ivezi{\'c}}, {\v Z}., {Jain}, A., {Jenness}, T., {Kanarek}, G., {Kendrew},
  S., {Kern}, N.~S., {Kerzendorf}, W.~E., {Khvalko}, A., {King}, J., {Kirkby},
  D., {Kulkarni}, A.~M., {Kumar}, A., {Lee}, A., {Lenz}, D., {Littlefair},
  S.~P., {Ma}, Z., {Macleod}, D.~M., {Mastropietro}, M., {McCully}, C.,
  {Montagnac}, S., {Morris}, B.~M., {Mueller}, M., {Mumford}, S.~J., {Muna},
  D., {Murphy}, N.~A., {Nelson}, S., {Nguyen}, G.~H., {Ninan}, J.~P.,
  {N{\"o}the}, M., {Ogaz}, S., {Oh}, S., {Parejko}, J.~K., {Parley}, N.,
  {Pascual}, S., {Patil}, R., {Patil}, A.~A., {Plunkett}, A.~L., {Prochaska},
  J.~X., {Rastogi}, T., {Reddy Janga}, V., {Sabater}, J., {Sakurikar}, P.,
  {Seifert}, M., {Sherbert}, L.~E., {Sherwood-Taylor}, H., {Shih}, A.~Y.,
  {Sick}, J., {Silbiger}, M.~T., {Singanamalla}, S., {Singer}, L.~P., {Sladen},
  P.~H., {Sooley}, K.~A., {Sornarajah}, S., {Streicher}, O., {Teuben}, P.,
  {Thomas}, S.~W., {Tremblay}, G.~R., {Turner}, J.~E.~H., {Terr{\'o}n}, V.,
  {van Kerkwijk}, M.~H., {de la Vega}, A., {Watkins}, L.~L., {Weaver}, B.~A.,
  {Whitmore}, J.~B., {Woillez}, J., {Zabalza}, V. and {Astropy Contributors}
  (2018) {The Astropy Project: Building an Open-science Project and Status of
  the v2.0 Core Package}.
\newblock AJ, 156, 123.
\newblock \url{https://doi.org/10.3847/1538-3881/aabc4f}.

\bibitem{oke95}
{Oke}, J.~B., {Cohen}, J.~G., {Carr}, M., {Cromer}, J., {Dingizian}, A.,
  {Harris}, F.~H., {Labrecque}, S., {Lucinio}, R., {Schaal}, W., {Epps}, H. and
  {Miller}, J. (1995) {The Keck Low-Resolution Imaging Spectrometer}.
\newblock PASP, 107, 375.
\newblock \url{https://doi.org/10.1086/133562}.

\bibitem{prz11}
{Przybilla}, N., {Nieva}, M.-F. and {Butler}, K. (2011) {Testing common
  classical LTE and NLTE model atmosphere and line-formation codes for
  quantitative spectroscopy of early-type stars}.
\newblock In Journal of Physics Conference Series, vol. 328 of \emph{Journal of
  Physics Conference Series}, p. 012015.
\newblock \url{https://doi.org/10.1088/1742-6596/328/1/012015}.

\bibitem{irr21}
{Irrgang}, A., {Geier}, S., {Heber}, U., {Kupfer}, T., {El-Badry}, K. and
  {Bloemen}, S. (2021) {A proto-helium white dwarf stripped by a substellar
  companion via common-envelope ejection. Uncovering the true nature of a
  candidate hypervelocity B-type star}.
\newblock A\&A, 650, A102.
\newblock \url{https://doi.org/10.1051/0004-6361/202038757}.

\bibitem{hub17}
{Hubeny}, I. and {Lanz}, T. (2017) {A brief introductory guide to TLUSTY and
  SYNSPEC}.
\newblock arXiv e-prints, arXiv:1706.01859.

\bibitem{dor22}
{Dorsch}, M., {Reindl}, N., {Pelisoli}, I., {Heber}, U., {Geier}, S.,
  {Istrate}, A.~G. and {Justham}, S. (2022) {Discovery of a highly magnetic
  He-sdO star from a double-degenerate binary merger}.
\newblock A\&A, 658, L9.
\newblock \url{https://doi.org/10.1051/0004-6361/202142880}.

\bibitem{hir09}
{Hirsch}, H.~A. (2009) {Hot subluminous stars : on the search for chemical
  signatures of their genesis}.
\newblock Ph.D. thesis, Friedrich Alexander University of Erlangen-Nuremberg,
  Germany.

\bibitem{cop10}
{Copperwheat}, C.~M., {Marsh}, T.~R., {Dhillon}, V.~S., {Littlefair}, S.~P.,
  {Hickman}, R., {G{\"a}nsicke}, B.~T. and {Southworth}, J. (2010) {Physical
  properties of IP Pegasi: an eclipsing dwarf nova with an unusually cool white
  dwarf}.
\newblock MNRAS, 402, 1824--1840.
\newblock \url{https://doi.org/10.1111/j.1365-2966.2009.16010.x}.

\end{thebibliography}

\end{document}